\documentclass[prb,twocolumn,preprintnumbers,amsmath,amssymb]{revtex4}

\usepackage{amsmath,latexsym,epsf,epsfig}

\newcommand{\II}{\mbox{${\mathbb I}$}}

\newcommand{\rd}{{\rm d}}
\newcommand{\diag}{{\rm diag}}
\newcommand{\D}{{\cal D}}
\newcommand{\U}{{\cal U}}

\newcommand{\ph}{\varphi}

\newcommand{\phl}{\varphi_{L}}
\newcommand{\phr}{\varphi_{R}}
\newcommand{\phz}{\varphi_{Z}}
\newcommand{\mur}{\mu_{{}_R}}
\newcommand{\mul}{\mu_{{}_L}}

\def\phz{\varphi_{Z}}

\def\a{\alpha}
\def\G{\mathbb G}
\def\UU{\mathbb U}
\def\S{\mathbb S}
\def\der{\partial }
\def\mis{{\frac{\rd k}{2\pi} }}
\def\ri{{\rm i}}
\def\xt{{\widetilde x}}

\def\e{{\rm e}}
\def\eps{\epsilon}
\def\k{\kappa}

\begin{document}
\title{Junctions of anyonic Luttinger wires}
\author{Brando Bellazzini$^1$, Pasquale Calabrese$^2$, and Mihail Mintchev$^2$}
\affiliation{
${}^1$ Scuola Normale Superiore and INFN, Pisa, Italy\\
${}^2$ Dipartimento di Fisica dell'Universit\`a di Pisa and INFN, Pisa, Italy}

\date{\today}

\begin{abstract}

We present an extended study of anyonic Luttinger liquids wires jointing at a
single point. 
The model on the full line is solved with bosonization and the junction of an
arbitrary number of wires is treated imposing boundary conditions that preserve
exact solvability in the bosonic language.
This allows us to reach, in the low momentum regime, some of the critical 
fixed points found with the electronic boundary conditions.
The stability of all the fixed points is discussed.

\end{abstract}

\maketitle

\section{Introduction}

\begin{figure}[t]
\includegraphics[width=\columnwidth]{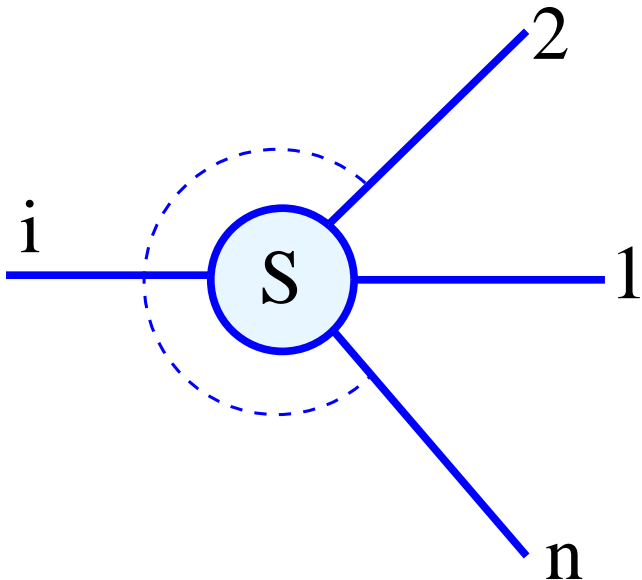} 
\caption{(Color online) 
A quantum junction of $n$ wires connected via a scattering matrix $\S$.}
\label{figjun}
\end{figure}

In the last few years there has been a boom in the study of transport 
properties at the junction of multiple quantum wires \cite{nfll-99,sdm-01,mw-02,y-02,lrs-02,cte-02,etcs-03,ppil-03,coa-03,rs-04,kvf-04,klvf-05,f-05,gs-05,emabms-05,kd-05,ff-05,drs-06,gw-06,bm-06,bms-07,bbms-08,hc-08,drs-08,dr-08,hkc-08}. 
This interest is largely motivated by the fact that junctions of three or
more wires would naturally appear in any quantum circuit.
Different frameworks have been developed to tackle this complicated 
problem that shows a rich phase diagram. 
In fact, despite of the universality in the bulk of the wires, that are 
described by a Luttinger liquid \cite{bos}, different conditions at 
the junctions can lead to exotic phase diagrams (as e.g. those 
in Refs. \onlinecite{nfll-99,coa-03,hc-08}) 
whose degree of universality is not yet understood.
According to the Renormalization Group (RG) theory of critical phenomena, 
the low energy properties of a gapless system are captured by the stable fixed
point of the RG flow, independently of microscopic (non-universal) details of
the real system. 
In view of the universality it is worthy to investigate very simple models, 
even exactly solvable, that can have (because of symmetry reasons) the same 
fixed points of the real systems. 
For bulk one-dimensional (1D) models and in the case of a single boundary,
conformal field theory provides a complete classification of the 
universality classes (see e.g. Ref. \onlinecite{cft}), 
whose analogous for junctions (or star-graph) is not yet known.
For all these reasons, we investigate in this paper the 
Tomonaga-Luttinger (TL) 
model on a junction with an arbitrary number $n$ of arms as depicted in Fig. 
\ref{figjun} (a junction with two wires $n=2$ can be seen as a defect on the
line, a problem that has been largely 
investigated \cite{kf-92,fn-93,wa-94,dms-94,oa-97,lm-98,s-98,kl-99,ll-99,mrs-01,Bachas:2001vj,ms-05}
in the past). 
To solve this problem, at the junction we impose conditions that are probably 
not obvious for an electronic problem, but they show the advantage to be 
exactly solvable. The natural hope is that the electronic model, at least for
some values of the couplings, would be in the domain of attraction of  
the fixed points found here.

Furthermore we calculate the transport for particles with generalized anyonic
statistics \cite{anyons}. The reason for this generalization is twofold. 
On one hand the study of 1D anyonic model is attracting a renewed interest 
\cite{k-99,bgo-06,oae-99,g-06,ssc-07,zw-92,cm-07,pka-07,an-07,lm-99,it-99,fibo,g-07,zw-07,sc-08,hzc-08,dc-08,bgk-08}, 
mainly motivated by possible experiments with cold atoms \cite{par}. 
On the other hand, the transport of wires joined with a quantum Hall
island is driven by anyonic excitations \cite{klvf-05}. 
Also in this case we can wonder whether the different problems have some
common fixed points.
In 1D, anyonic statistics are described in terms of fields that 
at different points ($x_1 \neq x_2$) satisfy the commutation relations
\begin{eqnarray}
\Psi^\dag(t,x_1) \Psi(t,x_2) &=& 
\e^{-\ri \pi \k \eps(x_{12})} \Psi(t,x_2)\Psi^\dag(t,x_1)\, , 
\nonumber\\
\Psi^\dag(t,x_1) \Psi^\dag(t,x_2) &=& 
\e^{\ri \pi \kappa \eps(x_{12})} \Psi^\dag(t,x_2)\Psi^\dag(t,x_1)\, ,
\label{anyon}
\end{eqnarray}
where $\eps(x)$ is the sign function [$\eps(z)=-\eps(-z)=1$ for $z>0$ and
$\eps(0)=0$] and $x_{12}=x_1-x_2$. 
$\k$ is called statistical parameter and equals $0$ for bosons and $1$ for 
fermions.
Other values of $\k$ give rise to general anyonic statistics ``interpolating''
between the two familiar ones.

The TL model emerges naturally in the description of spinless fermions in 1D
(and so electrons when the spin degrees of freedom are not
important, but spin is also easily introduced in the formalism). 
In fact, starting from fermions hopping on a chain, linearizing
the dispersion relation close to the Fermi surface at $\pm k_F$ and taking the
continuum limit, one arrives to the standard TL Hamiltonian \cite{bos} 
\begin{equation} 
{\cal H} = \int {dx}\left[ 
v_F (\psi_1^*\ri\der_x \psi_1 - \psi_2^*\ri\der_x\psi_2)  
+g_+\rho_+^2 
+ g_-\rho_-^2
\right]\, ,  
\label{lagrangian}
\end{equation} 
where $\psi_{1,2}(t,x)$ are the two complex fields representing 
free-fermions left
and right movers, $v_F$ is the Fermi velocity, i.e. the speed of the non 
interacting fermions, and 
\begin{equation}
\rho_\pm (t,x) = \left [\psi_1^*(t,x)\psi_1(t,x) \pm \psi_2^*(t,x)\psi_2(t,x) 
\right ]\, ,
\label{chargedensities}
\end{equation}
are the two independent charge densities.
All the interaction is encoded in the coupling constants   
$g_\pm$ [often the couplings $g_{2,4}=2(g_{+}\mp g_{-})$ are used].
Eventual irrelevant coupling terms of degree greater than four have been 
dropped. For $g_+>g_-$ the model is repulsive and it is attractive in the
opposite case.

A similar reasoning can be repeated for anyonic degrees of freedom and the
Hamiltonian is always given by Eq. (\ref{lagrangian}), but 
with $\psi_\a$ satisfying the commutation relations (\ref{anyon}),
$\psi_1$ with $\k$ and $\psi_2$ with $-\k$.
Thus, when $\k=1$ the model is the well-known fermionic TL model, while the 
bosonic limit $\k\to 0$ is not well defined in this formalism as will be
clearer in the following. 
We stress that this anyonic model is different from the gases discussed
elsewhere \cite{k-99,bgo-06,ssc-07,pka-07,sc-08}, 
that also have a Luttinger liquid description.  
As in the fermionic case, the model is naturally solved exactly 
through bosonization \cite{bos}.

This Hamiltonian defines completely the system on each wire. 
To complete the description of the junction like the one shown in 
Fig. \ref{figjun} we have to define the interaction between the $n$ wires.
From an electronic point of view it is natural to have a term of the 
form \cite{nfll-99}
\begin{equation} 
\psi^*_\a (t,0,i)\mathcal{B}_{\a \beta\; ij}\, \psi_\beta (t,0,j) \, , 
\label{blagrangian}
\end{equation} 
where $\a,\beta=1,2$ and $i,j=1\dots n$. The matrix $\mathcal{B}$ defines 
the boundary interaction among the fields $\psi$. 
Although very natural, this boundary condition is quite complicated after
bosonization, because it involves exponential boundary 
interactions of the bosonic fields. As a consequence the theory with this
interaction term is no longer exactly solvable with bosonization, and very
smart and complicated methods must be employed to extract the low-energy
behavior from it \cite{nfll-99,coa-03}. 
In this paper we take an alternative approach that is to modify the junction
couplings in such a way to preserve the exact solvability 
after bosonization \cite{bm-06,bms-07,bbms-08}. 
The main idea is to impose the boundary condition directly on the 
bosonic degrees of freedom trying to have the same symmetries as in the 
$\psi$ counterpart. 
The two problems can obviously have a different
structure of fixed points, but, as stressed above, the natural hope is that
the junction defined by Eq. (\ref{blagrangian}) shares some of the anyonic 
fixed points with the ones found here, as it is well known to happen for
fermions. The clear advantage of our approach is that keeping exactly
solvability, the results are obtained with a relative little effort, compared
to analogous ones for Eq. (\ref{blagrangian}).

The paper is organized as follows. 
In Sec. \ref{SecTL} we introduce the anyonic TL model and solve it on the full
line. In Sec. \ref{SecJ} after introducing the general features of the
junction and the importance of conservation laws, we first present the standard
solution on half-line and then generalize it to the generic junction.
In Sec. \ref{SecF} we study the stability of the found fixed points following
the RG flow. 
Finally in Sec. \ref{SecC} we draw our conclusion and discuss issues that
need further investigation.
In two Appendices \ref{appA} and \ref{appcrit} we report the technicality of
bosonization and the description of the fixed points of the junction. 

\section{The anyonic Tomonaga-Luttinger model}
\label{SecTL}

As already mentioned, the main goal of this paper is to investigate 
the Tomonaga-Luttinger model on half infinite quantum wires jointing in a 
single junction.  
However, in order to fix the notations and some basic tools, it is instructive 
to sketch first the solution of model on the line. In doing that 
we will focus on the general anyonic solution, which contains the 
more familiar fermionic one as a special case. 
The model is defined by the Hamiltonian (\ref{lagrangian}) in which the space
variable $x$ is integrated on the full real axis. 
The corresponding equations of motion are
\begin{multline}
\ri (\der_t -v_F\der_x) \psi_1(t,x)=\\
2 g_+\, \rho_+(t,x) \psi_1(t,x) + 2 g_-\, \rho_-(t,x) \psi_1(t,x)\,,\\
\shoveleft{\ri (\der_t +v_F\der_x) \psi_2(t,x)=}\\
2 g_+\, \rho_+(t,x) \psi_2(t,x)  + 2 g_-\, \rho_-(t,x) \psi_2(t,x)\,.
\label{eqm1}
\end{multline}
Bosonization \cite{bos} is the basic tool to quantize and solve 
these equations of motion. In fact, the solution can be expressed in terms of 
the right and left-moving scalar fields $\varphi_{R,L}$. 
The standard details of the solution can be found in textbooks \cite{bos}
and are reported in appendix \ref{appA} to make this paper self contained.  
The method is based on the change of variable
\begin{eqnarray}  
\psi_1(t,x) &\propto&
:\e^{\ri \sqrt {\pi} \left [\sigma \phr (vt-x) + \tau \phl (vt+x)\right ]}:\,, 
\label{psi1}\\
\psi_2(t,x) &\propto&
:\e^{\ri \sqrt {\pi} \left [\tau \phr (vt-x) + \sigma \phl (vt+x)\right ]}:\,, 
\label{psi2}
\end{eqnarray}
where the proportionality constants are explicitly given in appendix
\ref{appA},
and $: \cdots :$ denotes the normal product relative to the creation and 
annihilation operators of $\ph$ fields. 
$\sigma$, $\tau$ and $v$ are three real parameters to be determined
inserting these expressions in the equations of motion. 
Without loss of generality we take $\sigma \geq 0$ and assume that 
\begin{equation}
\sigma \not= \pm \tau \, .
\label{cond1}
\end{equation}
The charge densities take the very simple form
\begin{equation}
\rho_\pm (t,x) = \frac{-1}{2\sqrt {\pi }(\tau \pm \sigma)} 
\left [(\der \phr)(vt-x) \pm (\der \phl)(vt+x)\right ] . 
\label{rhopm}
\end{equation} 
Imposing the current conservation
\begin{equation}
\der_t \rho_\pm(t,x) -v\der_x j_\pm(t,x) = 0\, , 
\label{conservation}
\end{equation} 
one gets the currents
\begin{equation}
j_\pm (t,x) = \frac{(\der \phr)(vt-x) \mp (\der \phl)(vt+x)}{
2\sqrt {\pi} (\tau \pm \sigma)}\,.
\label{jpm}
\end{equation}  

Using the exchange properties of $\varphi$, one can easily show that 
that the field $\psi_1$ satisfies the anyonic commutation relations 
given in Eq. (\ref{anyon}) with statistical parameter
\begin{equation}
\kappa = \tau^2 - \sigma^2 \, 
\label{anyon3}
\end{equation}
According to Eq. (\ref{cond1}) $\kappa \not= 0$, that shows explicitly that the
bosonic limit is not well defined in this context. 
The exchange relations of $\psi_2$ follow from Eq. (\ref{anyon}) with the 
substitution $\kappa \mapsto -\kappa$, implying that $\psi_\a$ are both 
anyon fields, which become canonical fermions for $\kappa =1$. 

The quantum equations of motion are obtained from Eq. (\ref{eqm1}) by 
replacing 
$\rho_\pm(t,x) \psi_\a (t,x) \longmapsto :\rho_\pm \psi_\a:$, 
giving 
\begin{eqnarray}
\tau (v-v_F)\pi = \frac{g_+}{\tau + \sigma} +\frac {g_-}{\tau-\sigma}\, , 
\label{qeqm1}\\
\sigma (v+v_F)\pi = \frac{g_+}{\tau + \sigma} -\frac {g_-}{\tau-\sigma}\, ,
\label{qeqm2}
\end{eqnarray} 
which combined with Eq. (\ref{anyon3}) determine $\sigma$, $\tau$ and the 
velocity $v$ in terms of the coupling constants $g_\pm$ and the 
statistical parameter $\kappa$. 
In terms of the variables $\zeta_\pm =\tau\pm\sigma$, one obtains the system
of equations 
\begin{eqnarray}
\zeta_+\, \zeta_-&=&\kappa\, , 
\label{sys1} \\
v \zeta_+^2 &=& v_F \kappa +\frac{2}{\pi}g_+ \, , 
\label{sys2}\\
v \zeta_-^2 &=& v_F \kappa +\frac{2}{\pi}g_- \, ,  
\end{eqnarray}
with solution 
\begin{eqnarray}
\zeta_\pm^2 &=& |\kappa|
\left(\frac{\pi \kappa v_F+2g_+}{\pi \kappa v_F+2g_-}\right)^{\pm 1/2}\, , 
\label{z}\\ 
v&=&\frac{\sqrt{(\pi \kappa v_F+2g_-)(\pi \kappa v_F+2g_+)}}{\pi|\kappa|}\,. 
\label{v}
\end{eqnarray}
The relations (\ref{z}) and (\ref{v}) are the anyonic realization 
of the well known result valid for canonical fermions in the TL 
model (the traditionally used parameter $K$ in our notation
coincides for $\kappa=1$ with $\zeta_{-}^2=\zeta_+^{-2}$, for comparison in 
Refs. \onlinecite{coa-03,hc-08} the notation is $g=K^{-1}$). 
The stability conditions of the model is $2g_\pm> -\pi \kappa v_F$ that 
ensures $\sigma$, $\tau$ and $v$ to be real and finite.

From the previously given mapping it is easy to write the Hamiltonian in 
terms of the bosonic fields, obtaining
\begin{equation}
{\cal H}=\frac{v}2\int dx 
\left[(\der_x \theta)^2+(\der_x \ph)^2\right]\,,
\label{act1}
\end{equation} 
where 
\begin{eqnarray} 
\ph (t,x) = \frac{1}{2} \left [\ph_R(vt-x) + \ph_L(vt+x)\right ] \,, 
\label{ph}\\
\theta (t,x) = \frac{1}{2} \left [\ph_R(vt-x) - \ph_L(vt+x)\right ] \,, 
\end{eqnarray} 
where $\theta$ is the so-called dual field. 
Notice that the Hamiltonian is slightly different from the usual one
in the literature because we adsorb the coupling constant $g$ (or $K$) 
in the definition of the fields.

It is worth commenting at this point the internal symmetries of the TL
Hamiltonian,  
because they will characterize the quantization on the junction.
The TL Hamiltonian (\ref{lagrangian}) is left invariant by the two independent 
$U(1)$ phase transformations usually denoted 
as $U(1)\otimes {\widetilde U}(1)$:
\begin{eqnarray}
\psi_\a &\rightarrow& \e^{\ri s} \psi_\a \, , \quad \qquad \; \, 
\psi^*_\a \rightarrow \e^{-\ri s} \psi^*_\a\, ,  
\label{V}\\
\psi_\a &\rightarrow& \e^{-\ri(-1)^\a {\tilde s}} \psi_\a\,,\quad 
\psi^*_\a \rightarrow \e^{\ri(-1)^\a {\tilde s}} \psi^*_\a\, . 
\label{A}
\end{eqnarray} 
In the bosonic language they correspond to the independent 
shift-invariance of the (compactified) fields $\ph_{R,L}$. 
We will see that on the junction, the left and right movers are not 
independent anymore and the
two $U(1)$ symmetries cannot be conserved simultaneously.

One of the main advantages of bosonization is that after having solved the
equations of motion, it is straightforward to obtain all the correlation 
functions (also at finite temperature) just by commuting the fields $\ph$ 
in the exponential forms for $\psi$, using Eq. (\ref{identity1}). 
In fact, in terms of the basic correlator
\begin{equation}
{\cal D} (x)=\frac1{\ri(x-\ri\eps)}\,,
\label{DD}
\end{equation}
the zero-temperature (Fock representation) field correlation functions are
\begin{multline}
\langle \psi_1^* (t_1,x_1) \psi_1 (t_2,x_2) \rangle  \\ 
=\frac1{2\pi} 
[{\cal D} (vt_{12}-x_{12})]^{\sigma^2}[{\cal D}(vt_{12}+x_{12})]^{\tau^2}\,,\\
\shoveleft{\langle \psi_2^* (t_1,x_1) \psi_2 (t_2,x_2) \rangle} \\
=\frac1{2\pi} 
[{\cal D} (vt_{12}-x_{12})]^{\tau^2}[{\cal D}(vt_{12}+x_{12})]^{\sigma^2}\,,
\end{multline}
with $x_{12}=x_1-x_2$ and $t_{12}=t_1-t_2$.
Scale invariance is manifest and one can read the dimension of $\psi_\a$ 
\begin{equation}
d_{{\rm line}}=
\frac{1}{2}(\sigma^2 + \tau^2)=\frac{1}{4}(\zeta_{+}^{2}+\zeta_{-}^{2})\, . 
\label{linedim}
\end{equation} 
All the other two-point field correlation functions vanish because 
of Eq. (\ref{cond1}) and the neutrality 
condition ($U(1)\otimes {\widetilde U}(1)$-symmetry).
Analogously for the $U(1)$-density one finds
\begin{multline}
\langle \rho_+ (t_1,x_1) \rho_+ (t_2,x_2) \rangle  \\= 
\frac{1}{(2\pi\zeta_+)^2} 
\left [[{\cal D} (vt_{12}-x_{12})]^2+[{\cal D} (vt_{12}+x_{12})]^2
\right ]\, , 
\label{cf5}
\end{multline}
and straightforwardly the ones for $\rho_-$ and $j_\pm$ are obtained.
We notice that all these correlation functions correctly agree with the
general expression for an harmonic anyonic fluid\cite{cm-07} with only one
harmonic term given by the Luttinger mode.

The generalization to finite temperature $\beta^{-1}$ (Gibbs representation)
is simply obtained with the replacement ${\cal D}(x)\to {\cal D}_\beta(x)$
with
\begin{equation}
{\cal D}_\beta(x)= \left[\frac{\ri \beta}{\pi}\sinh \left ( \frac{\pi x}{\beta} -\ri \eps \right )\right ]^{-1}\,,
\end{equation} 
and introducing the chemical potentials, explicitly
\begin{multline}
\langle \psi_1^* (t_1,x_1) \psi_1 (t_2,x_2) \rangle _\beta \\= 
\frac{1}{2\pi} 
\e^{\ri \mur \sigma (vt_{12}-x_{12}) + \ri \mul \tau (vt_{12}+x_{12})} \\
\times
[{\cal D}_\beta (vt_{12}-x_{12})]^{\sigma^2}[{\cal D}_\beta (vt_{12}+x_{12})]^{\tau^2}\,,
\end{multline}
and similarly for the other correlations.
The right and left chemical potentials are 
\begin{equation} 
\mur = \frac{\mu}{\zeta_+}  - \frac{\tilde \mu}{\zeta_-}\, , \quad 
\mul = \frac{\mu}{\zeta_+} + \frac{\tilde \mu}{\zeta_-}\, ,
\label{cpotentials} 
\end{equation}
where $\mu$ and ${\tilde \mu}$ are the ones associated with the 
$U(1)\otimes {\widetilde U}(1)$-charges.

\section{The Junction of Tomonaga-Luttinger liquids} 
\label{SecJ}

\subsection{Boundary conditions and symmetries}

After the previous preliminary considerations on the line, we investigate below
the TL model at a junction like the one shown in Fig. \ref{figjun}.
In mathematical physics literature these junctions are usually called 
{\it star graphs} and they represent the building blocks for more 
general ``quantum graph'' networks (see for a review Ref. \onlinecite{qg}). 
We now fix all the notation on the junction that we call $\Gamma$. 
We indicate the jointing point of the junction as $V$.
Each point $P$ in the bulk $\Gamma \setminus V$ (i.e. of the wires)
can be parametrized by the pair $(x,i)$,
where $i=1,\ldots,n$ labels the edge $E_i$ and $x\in(0,\infty)$ 
is the distance of $P$ from the vertex $V$ along that edge. 
We stress that, as physically suggested, the embedding of $\Gamma$ and the 
relative position of the edges in the ``ambient space'' are irrelevant.

The dynamics of each wire (edge) is still given by the Hamiltonian
(\ref{lagrangian}), but now $\psi_\a = \psi_\a (t,x,i)$ 
and $x>0$. 
As already discussed in the introduction, in order to fix the solution one 
must impose some boundary conditions at the vertex $V$ at $x=0$. 
The simplest boundary condition one can imagine is linear in $\psi_\a$ 
and is generated by the boundary term in Eq. (\ref{blagrangian}) that 
makes the model non-exactly solvable for general couplings
(see e.g. Ref. \onlinecite{nfll-99} for free fermions and also 
\onlinecite{coa-03} for infinite repulsive coupling).

An alternative which preserves the exact solvability after bosonization has 
been proposed \cite{bm-06,bms-07,bbms-08}. 
The main idea is to impose the boundary condition directly on the 
bosonic degrees of freedom, selecting those of them which ensure unitary time 
evolution of the fields $\varphi$. 
This is guaranteed only if the boundary conditions are linear in the fields
$\ph$ and its first derivatives. So we can parametrize 
these boundary conditions by a generic $n\times n$ unitary 
matrix $\UU$\cite{ks-00,h-00,bm-06,bms-07} 
\begin{equation} 
\sum_{j=1}^n \left [\lambda (\II -\UU)_{ij}\, \ph (t,0,j) 
-\ri (\II+\UU)_{ij} (\partial_x\ph ) (t,0,j)\right ] = 0\, , 
\label{bc} 
\end{equation} 
and $\lambda >0$ is a parameter with dimension of mass needed to recover 
the correct physical dimensions.
Since bosonization expresses physical charges linearly in $\varphi$, we shall 
see below that these boundary conditions 
simply state how the charges are parcelled out among the wires at the vertex.

The analysis of the fixed point is greatly simplified if we assume 
time-reversal invariance. This implies that the matrix $\UU$ must be real, 
that together with unitarity leads to a symmetric matrix $\UU$, i.e. 
\begin{equation} 
\UU^t = \UU\, ,  
\label{tr}
\end{equation}
giving a further constraint on the possible boundary terms.
A non trivial magnetic flux (breaking time-reversal) has been 
considered\cite{coa-03} and resulted in a more complicated fixed point 
structure. When dealing with anyon excitation, it would be more natural to
consider non time-reversal models, because the magnetic field needed
to produce the anyons breaks the symmetry.
However this would complicate the analysis and in some regime it could
be only an irrelevant perturbation. Thus in the following we will always
assume time-reversal invariance and leave the study of the effect of its
breaking to a future work.

The boundary condition (\ref{bc}) is equivalent \cite{bm-06,bms-07} to an 
interaction with a point-like defect localized at the vertex of the graph. 
The scattering matrix associated with this 
interaction is \cite{ks-00,bm-06,bms-07}
\begin{equation} 
\S(k) = -[\lambda(\II-\UU)+k(\II+\UU)]^{-1} [\lambda (\II-\UU) - k(\II+\UU)], 
\label{S1}
\end{equation} 
and has transparent physical meaning: the diagonal element $\S_{ii}(k)$ 
represents the reflection amplitude on the edge $E_i$, 
whereas $\S_{ij}(k)$ with $i\not=j$ equals the transmission amplitude from 
$E_i$ to $E_j$. 
Eq. (\ref{S1}) makes also clear the meaning of the boundary terms $\lambda$
and $\UU$: for $\lambda\neq 0$ we have $\S(k=\lambda)=\UU$, i.e. $\lambda$ 
fixes
the momentum scale at which the scattering matrix is given exactly by $\UU$.

By construction the scattering matrix (\ref{S1}) is unitary 
\begin{equation}  
[\S(k)]^*=[\S(k)]^{-1} \, , 
\label{unit1}
\end{equation}  
and satisfies Hermitian analyticity 
\begin{equation}  
[\S(k)]^*=\S(-k)\, .  
\label{Ha}
\end{equation}  
Moreover, time reversal invariance (\ref{tr}) implies 
\begin{equation}  
[\S(k)]^t = \S(k)\, . 
\label{tr2} 
\end{equation} 
For simplicity we assume in this paper that $\UU$ is such that 
\begin{equation}
\int_{-\infty}^{\infty} \frac{\rd k}{2\pi } \e^{\ri kx} \S_{ij} (k) = 0\, , 
\qquad x>0\, , 
\label{bs}
\end{equation}
which guarantees that $\S(k)$ has no bound states (see \onlinecite{bs} 
for an extension to bound states).

The boundary conditions strongly influence the symmetry content on the
junction. 
Each symmetry in the bulk gives a conserved charge $Q$ [with density
$\rho(x,t)$] because of the Noether theorem. 
If we want to keep the conservation of $Q$ at the junction we {\it must}
impose from the beginning that the currents $j(x,t)$ corresponding to the 
given density $\rho(x,t)$ are conserved at the vertex.
This results in $\sum_{i=1}^n j (0,t)=0$ for all times.  
This is the {\it Kirchhoff's rule}, which must be imposed in the vertex  
in order to generate a time-independent charge from a given current. 
A basic example is given by the energy, that is a conserved quantity in the
bulk. Because of unitarity, the matrix $\UU$ in Eq. (\ref{bc}) parametrizes all 
boundary conditions which ensure the Kirchhoff rule for the energy-momentum 
tensor of $\ph$ and thus the time-independence of the relative Hamiltonian.
This means that there is no dissipation at the junction: if the energy flows 
out from one wire should flow in another one.
We stress that the Kirchhoff's rule for gapless models on a graph 
is the generalization of the celebrated result that scale invariance implies
holomorphic and antiholomorphic components of the energy tensor to be equal in
boundary conformal field theory \cite{c-84,cft}.  

Energy is not the only conserved quantity. In our formalism it is
conserved by construction, but all other conservation laws we want to keep on
the junction must be imposed by hand with appropriate Kirchhoff's rules.
However it may happen that different conserved currents can generate 
contradictory Kirchhoff's rules, resulting in obstructions for 
lifting all symmetries on the line to symmetries on $\Gamma$ \cite{foot}. 
In this case one  can preserve on $\Gamma$ one of the corresponding
symmetries, but not all of them. 
This is actually the case for the $U(1)\otimes{\widetilde U}(1)$-symmetry of
the TL model. 
In fact, the relative Kirchhoff rules generate \cite{bms-07} 
the following further constraints on $\UU$ 
\begin{eqnarray} 
\sum_{i=1}^{n}j_{+}(t,0,i)=0 \Longleftrightarrow \sum_{i=1}^{n}\S_{ji}(k)=
\sum_{i=1}^{n}\UU_{ji}=1\, , 
\label{kirchV}\\
\sum_{i=1}^{n}j_{-}(t,0,i)=0\Longleftrightarrow \sum_{i=1}^{n}\S_{ji}(k)=
\sum_{i=1}^{n}\UU_{ji}=-1\, ,
\label{kirchA}
\end{eqnarray}
which cannot be satisfied simultaneously. 
$U(1)$ is linked to the electric charge conservation and it is then natural to
require the conservation of Eq. (\ref{kirchV}), while breaking  
Eq. (\ref{kirchA}). However also the opposite prescription has some interest.
Notice that the duality transformation (\ref{dual1}) on $\Gamma$ maps the 
matrix $\UU$ [and so $\S(k,\lambda)$] in $-\UU$ [$-\S(k^{-1},\lambda^{-1})$]. 
Consequently duality maps the
vertex conservation of $U(1)$ in ${\widetilde U}(1)$.

The matrix conductance $\G$ of the junction can be obtained in linear response 
theory. Since it involves only currents, the calculation is the same as
for free bosons \cite{bms-07,bbms-08}, but with the renormalized current
in Eq. (\ref{jpm}), leading to an overall normalization:
\begin{equation}
\G=\frac1{2\pi\zeta_+^2} (\II- \S)=G_{\rm line}(\II- \S) \,. 
\end{equation}
Thus the dependence of the conductance on the anyonic parameter is only 
through the renormalization constant $\zeta_+$ in Eq. (\ref{z}).
Because of unitarity $|\S_{ii}|\leq 1$, we have 
\begin{equation}
0\leq \G_{ii}\leq 2 G_{\rm line} \,.
\end{equation}
In the following, we will call conductance $G$ the diagonal element 
$\G_{ii}$ in the case it does not depend on the wire index $i$. 

It is worth mentioning that a similar approach (called Delayed Evaluation of 
Boundary Condition) working also with fermion boundary conditions has been 
developed by Chamon et al. \cite{coa-03,hc-08}.
It basically amounts to leave in the half-line, right and left movers 
unconstrained in the bulk, constructing then the tunneling operators, 
and only later choosing an $R$-matrix ($R$ for reflection, it can be easily 
rewritten as an $S$-matrix) such that one of these processes pins the 
correct boundary conditions.
In the appendix A of Ref. \onlinecite{hc-08} the conductance is written 
in terms of an $n\times n$ $R$, which agrees with the results here and 
elsewhere \cite{bms-07,bbms-08}. 

\subsection{The half-line} 

It is instructive to start with the well-known case $n=1$, 
namely the half-line, since 
some features of the generic junction are already manifest in this case. 
The matrices $\UU$ and $\S$ are just numbers $U$ and $S$.
Setting $U= \e^{-2\ri \a}$, we get 
\begin{equation} 
S(k) = \frac{k-\ri \eta}{k+\ri \eta} \, , 
\label{Shl}
\end{equation}
with  
\begin{equation}  
\eta=\lambda\tan(\a)\,,\qquad -\frac{\pi}{2}\leq\a\leq\frac{\pi}{2}\,.
\end{equation}
As expected the $S$-matrix (\ref{Shl}) corresponds to full reflection and 
describes the mixed (Robin) boundary condition 
\begin{equation} 
(\partial_x\ph ) (t,0) - \eta\, \ph (t,0)=0\, . 
\label{Robin}
\end{equation} 
The condition (\ref{bs}) implies $\eta \geq 0$ or equivalently 
$0\leq \a \leq \pi/2$.  
$\a=0$ and $\a =\pi/2$ correspond to Neumann and Dirichlet 
boundary conditions respectively. 
These two points define the {\it only} bosonic scale invariant boundary 
conditions on the half-line. 
Instead of imposing the condition (\ref{Robin}), we can add a term to the
Hamiltonian in such a way to generate it as a further equation of motion. 
The resulting total Hamiltonian is
\begin{equation}
{\cal H}_{\rm Tot}= {\cal H}+\eta \ph^2(t,0)\,,
\label{Hambou}
\end{equation}
with ${\cal H}$ the bulk term given by Eq. (\ref{act1}), obviously defined 
only on the half-line, i.e. the integral is over $x\in (0,\infty)$.

The main effect of the boundary in $x=0$ is to couple right and left 
movers by means of the boundary condition (\ref{Robin}). In particular, 
at criticality, Eq. (\ref{Robin}) implies that
\begin{eqnarray}
\phl(\xi)=\phr(\xi),&\qquad \eta=0,\\
\phl(\xi)=-\phr(\xi),&\qquad \eta=\infty,
\end{eqnarray}
which is the familiar ``unfolded picture''\cite{nfll-99} 
for Neumann and Dirichlet boundary conditions. 
The boundary conditions then forces non-zero mixed commutation relation [from
Eqs. (\ref{crl}) and (\ref{clr})] between right and left movers 
\begin{multline}
[\phr(\xi_{1}),\phl(\xi_{2})]=\\
\left\{
\begin{array}{ll}
-\ri\eps(\xi_{12})\,,& \eta=0\,,\\
\ri\eps(\xi_{12})\,, & \eta=\infty\,,\\
\ri\eps(\xi_{12})-4\ri\theta(\xi_{12})e^{-\eta\xi_{12}}, & 0<\eta<\infty,
\end{array}
\right.
\label{RLhl}
\end{multline}
while the left-left and right-right ones are the same as in the full line. 
Note that in the right-left commutators it appears 
$\xi_{12}=vt_{12}-\xt_{12}$, 
involving, as expected, the sum of distances from the boundary
$\xt_{12}=x_1+x_2$. 

Although right and left modes are no longer decoupled,
we can still perform the bosonization program and solve the TL model
on the half-line. 
The anyonic exchange relations (\ref{anyon}) are still valid 
defining $\psi_{\a}$ as in Eqs. (\ref{psi1}) and (\ref{psi2}) [but with  
normalization constants depending on the boundary conditions, see
Eq. (\ref{z2})]. 
$\psi_{\a}$ fulfills the quantum equations of motion of
the TL model restricted to the half-line $x>0$, 
with $\sigma$, $\tau$ and $v$ given by the same expressions 
(\ref{z}) and (\ref{v}) found for the full line. 
In fact, all the local bulk relations of the TL model on the full line still
hold on the half-line. 
This will remain true in the more general case of a junction made of
any number of wires. 

The charge and current densities Eqs. (\ref{rhopm}) and (\ref{jpm})   
are still locally conserved  [i.e. Eq. (\ref{conservation}) holds 
for $x\neq0$],
and $\rho_{\pm}$ generate the $U(1)\otimes\tilde{U}(1)$ infinitesimal 
transformations  (\ref{infinitesimal1}).
After bosonization, the boundary condition (\ref{Robin}) 
can be recasted in terms of physical currents
\begin{equation}
\begin{array}{ll}
j_{+}(t,0)=0\,, & \quad \eta=0\,,\\
j_{-}(t,0)=0\,, & \quad \eta=\infty\,,\label{bcj}\\
\partial_{x}j_{-}(t,0)-\eta j_{-}(t,0)=0\,, & \quad 0<\eta<\infty\,. 
\end{array}
\end{equation}
Consequently, the main physical difference between half and full line concerns 
the global charges $Q$ and $\widetilde{Q}$ associated  
to charge densities $\rho_{+}$ and $\rho_{-}$ respectively. 
The boundary spoils the simultaneous conservation of both charges, 
allowing just one linear combination to survive.
For instance, at the critical point $\eta=0$, the boundary 
condition (\ref{Robin}) is simply the Kirchhoff's rule associated to the 
$U(1)$ transformation (\ref{V}), enforcing the charge density current 
$j_{+}$ to vanish at the vertex while $j_{-}$ does not
\begin{equation}
j_{+}(t,0)=0\,,\qquad j_{-}(t,0)\neq0,\qquad {\rm for}\; \eta=0\,.
\end{equation}
In this case $Q$ is time-independent, while $\tilde{Q}$ depends on time 
due to a nontrivial charge flow through the boundary.
The critical point $\eta=\infty$ has an opposite behavior, 
preserving the $\tilde{U}(1)$ transformation (\ref{A}), 
and breaking (\ref{V}).
For generic finite $\eta>0$, it is easy to see that the $\tilde{U}(1)$
symmetry is always conserved while $U(1)$ is broken\cite{lm-98}.
As already pointed out, this symmetry breaking from 
$U(1)\otimes\tilde{U}(1)$ to a subgroup $U(1)$ 
is a general unavoidable feature of junctions  of any number of wires.

This boundary symmetry breaking is even more visible in the correlation 
functions. 
In addition to the usual right-right and left-left bosonic correlators,  
there are also mixed ones, Eqs. (\ref{LLRR}), (\ref{crl}), and (\ref{clr}).
As a consequence there are four non vanishing 2-points correlators 
for $\psi_{\a}$, instead of just two as for the full line.
For instance, considering the critical case $\eta=0$, when the $U(1)$
transformation (\ref{V}) is preserved, we have
\begin{multline}
\langle \psi_1^* (t_1,x_1) \psi_1 (t_2,x_2) \rangle=
  \langle \psi_1 (t_1,x_1) \psi_1^* (t_2,x_2) \rangle=\\
\left[{\cal D} (vt_{12}-x_{12})\right]^{\sigma^2}
\left[{\cal D}(vt_{12}+x_{12})\right]^{\tau^2} \\ \times
\left[{\cal D}(vt_{12}-\xt_{12}){\cal D}(vt_{12}+\xt_{12})\right]^{\sigma\tau},
\label{cfhl1}
\end{multline}
\begin{multline}
\langle \psi_1^* (t_1,x_1) \psi_2 (t_2,x_2) \rangle=
\langle \psi_2 (t_1,x_1) \psi_1^* (t_2,x_2) \rangle=\\
\label{cfhl2}
\left [{\cal D}(vt_{12}-x_{12})\right]^{\sigma \tau}
\left [{\cal D}(vt_{12}+x_{12})\right]^{\sigma \tau} \\
\times 
\left [{\cal D}(vt_{12}-\xt_{12})\right]^{\sigma^2}
\left [{\cal D}(vt_{12}+\xt_{12})\right]^{\tau^2 }\, , 
\end{multline}
and
\begin{eqnarray}  
\langle \psi_2^* (t_1,x_1) \psi_2 (t_2,x_2) \rangle  
= &\langle \psi_2 (t_1,x_1) \psi_2^* (t_2,x_2) \rangle\\
=&(\ref{cfhl1})\quad {\rm with}\quad  \sigma \leftrightarrow \tau \nonumber \\
\langle \psi_2^* (t_1,x_1) \psi_1 (t_2,x_2) \rangle  
= &\langle \psi_1 (t_1,x_1) \psi_2^* (t_2,x_2) \rangle 
\label{cfhl3}\\
=&(\ref{cfhl2})\quad {\rm with}\quad  \sigma \leftrightarrow \tau\,, \nonumber
\end{eqnarray} 
with $\xt_{12}=x_1+x_2$.
The non-triviality of the correlators (\ref{cfhl2}) and (\ref{cfhl3}) 
reflects the breaking of 
the $\tilde{U}$-symmetry on the half-line for $\eta=0$.

All the correlation functions just derived must be compared with the general 
scaling form
coming from boundary conformal field theory \cite{c-84} that in imaginary time
$\tau_i=\ri t_i$ predicts in general
\begin{equation}
\langle\Psi^*(z_1) \Psi(z_2) \rangle=
\left(\frac{1}{z_{12}z_{\bar1\bar2}}\right)^{d_{\rm line}}
F(\xi)\,,
\label{2ptgen}
\end{equation}
with the four point ratio
\begin{equation}
\xi=\frac{z_{1\bar1}z_{2\bar2}}{z_{1\bar2}z_{2\bar1}}\,,
\end{equation}
and $z_i=x_i+\ri\tau_i$, $z_{\bar i}={\bar z}_i$.
$F(\xi)$ encodes all the boundary dependence and for small argument can be
written as\cite{c-84} $F(\xi\ll 1)\propto \xi^{d_b}$, where $d_b$ is called
boundary exponent.
The real time correlations we wrote are clearly not of this form, but this is
just because we wrote them in the regime $x_1,x_2\gg 1$ and $x_{12}$,
$\xt_{12}$ arbitrary using the definitions (\ref{psi1}) and (\ref{psi2}). 
If we want to get the correct 
scaling also for arbitrary $x_{1,2}$ we should modify the definitions as 
\begin{eqnarray}  
\psi_1(t,x) &\propto&
:\e^{\ri \sqrt {\pi} \sigma \phr (vt-x)}::e^{\ri\sqrt{\pi}\tau\phl (vt+x)}:\,, 
\\
\psi_2(t,x) &\propto&
:\e^{\ri \sqrt {\pi} \tau\phr (vt-x)}: :e^{\ri\sqrt{\pi}\sigma\phl (vt+x)}:\,, 
\end{eqnarray}
at the price of introducing some more divergences that are easily
renormalized. With this prescription, we obtain as a typical example 
\begin{multline}
\langle \psi_1^* (t_1,x_1) \psi_1 (t_2,x_2) \rangle=
  \langle \psi_1 (t_1,x_1) \psi_1^* (t_2,x_2) \rangle=\\
\left[{\cal D} (vt_{12}-x_{12})\right]^{\sigma^2}
\left[{\cal D}(vt_{12}+x_{12})\right]^{\tau^2} \\ \times
\left[
\frac{{\cal D}(vt_{12}-\xt_{12}){\cal D}(vt_{12}+\xt_{12})}{
{\cal D}(2x_1){\cal D}(2x_2)}\right]^{\sigma\tau},
\label{bcf1}
\end{multline}
that agrees with the general conformal field theory scaling with 
$F(\xi)=\xi^{\sigma\tau}$ and so $d_b=\sigma\tau$. All the other correlation
functions are easily modified accordingly. 
Because it will be easier to write, in the following we will ignore the
double normal product and still use definitions 
(\ref{psi1}) and (\ref{psi2}). The expressions taking into account the correct
normalization at the boundary can be easily written down from the correlation 
we will derive.

We finally point out that for Dirichlet boundary conditions,
i.e. $\eta=\infty$, the diagonal correlations are the same but with
$d_b=-\sigma\tau$. Non diagonal correlations can be found in Ref. 
\onlinecite{lm-98}.

\subsection{Generic junction} 

The case of a junction with an arbitrary number $n>1$ of wires can be actually 
reduced to the study of a suitable family of $n$ half-lines. 
In fact, let $\U$ be the unitary matrix diagonalizing $\UU$ which defines the 
boundary conditions (\ref{bc}). 
Since $\UU$ is symmetric, we can choose $\U$ orthogonal, $\U^t=\U^{-1}$, 
and real, ${\U}^*=\U$. 
Let us parametrize the diagonal form
\begin{equation} 
\UU_d=\U\, \UU\, \U^{-1}
\label{d1}
\end{equation}  
as follows 
\begin{equation} 
\UU_d = \diag 
\left(\e^{-2\ri\a_1},\e^{-2\ri\a_2},\dots,\e^{-2\ri\a_n}\right)\,. 
\label{d2}
\end{equation}  
Using the definition (\ref{S1}) of $\S(k)$, one easily verifies that $\U$ 
diagonalizes $\S(k)$ for any $k$ and that 
\begin{multline} 
\S_d(k) = \U \S(k) \U^{-1} =\\ 
\diag \left (\frac{k-\ri \eta_1}{k+\ri \eta_1}, \frac{k-\ri \eta_2}{k+\ri \eta_2}, ... , \frac{k-\ri \eta_n}{k+\ri \eta_n} \right )\, , 
\label{d3}
\end{multline} 
where 
\begin{equation} 
\eta_i = \lambda \tan (\a_i)\, , \qquad -\frac{\pi}{2} \leq \a_i \leq \frac{\pi}{2}\, .  
\label{p1}
\end{equation}  
Therefore $\S(k)$ is a meromorphic function in the complex $k$-plane, whose 
poles are different from $0$ and are all located on the imaginary axis. 
The condition (\ref{bs}) 
implies absence of bound states i.e. of poles in the upper complex $k$-plane, 
namely $0\leq \a_i \leq \pi/2$, hence $\eta_{i}\ge0$.
 
Critical boundary conditions correspond to a matrix $\UU$ such that the 
scattering matrix is insensitive to the momentum scale transformations  
$\lambda\rightarrow\varrho\lambda$ (or $k\rightarrow \varrho^{-1}k$) 
with $\varrho>0$.
To be scale invariant, the scattering matrix must have each $\eta_{i}$ 
vanishing or infinite, so that $S$ is actually momentum independent 
and with eigenvalues $\pm1$. 
By means of Eqs. (\ref{unit1}), (\ref{Ha}), and the derivative\cite{bms-07}
\begin{equation}
k\frac{d \S(k)}{dk}=-\frac{1}{2}\left[\S(k)-\S^*(k)\right]\S(k)\,,
\label{Sflow}
\end{equation}
we see that criticality is equivalent to the condition 
\begin{equation}
\S=\S^*\,.
\label{crit}
\end{equation}
In appendix \ref{appcrit} some examples of critical junctions 
with two, three and four wires are given.

The matrix $\U$ allows us to define real scalar fields $\ph^d=\U\ph$ 
which are not localized on the single edges but have simple boundary 
conditions, formally the ones of disjoined half-lines
\begin{equation} 
(\partial_x\ph^d )(t,0,i)- \eta_{i}\, \ph^d (t,0,i)=0\,,
\quad i=1,\ldots,n\,.
\label{bchlg}
\end{equation} 
Comparing with the half-line Eqs. (\ref{RLhl}), it is straightforward to 
derive the commutation relations for the right and left movers on the wires
as done in Refs. \onlinecite{bm-06,bms-07,bbms-08} and reported in the
appendix \ref{appA}.

\subsection{The TL model at the junction}

The TL model on the star graph $\Gamma$ is defined by the sum of $n$
Hamiltonians in Eq. (\ref{lagrangian}) plus the boundary term that we implement
through Eq. (\ref{bc}) at the bosonic level. 
The charges on each wire are defined via Eq. (\ref{chargedensities})
and generate the $U(1)\otimes {\widetilde U}(1)$ phase transformations 
(\ref{infinitesimal1}) and (\ref{infinitesimal2}) 
leaving the Hamiltonian invariant.
The corresponding quantum equations of motion in the bulk 
are given by Eqs. (\ref{eqm1}) for each wire independently.
 
In analogy with Eqs. (\ref{psi1}) and (\ref{psi2}), the solution of the 
equations of motion is given by the vertex operator  
\begin{eqnarray}  
\psi_1(t,x,i) &\propto& 
:\e^{\ri \sqrt{\pi}\left [\sigma \ph_{i,R}(vt-x) + 
\tau \ph_{i,L}(vt+x)\right]}: \, , 
\label{psi1g}\\
\psi_2(t,x,i) &\propto& 
:\e^{\ri \sqrt{\pi}\left [\tau \ph_{i,R}(vt-x) + 
\sigma \ph_{i,L}(vt+x)\right]}: \, , 
\label{psi2g}
\end{eqnarray} 
where the normalization constants are given in the appendix \ref{appA} and 
depend on the anyon Klein factors. 
All bulk relations (the value of $\sigma$, $\tau$ and $v$, the form of the
currents etc.) of TL model on the line 
are still valid for half infinite wires jointed in a single vertex.

It is interesting to rewrite the boundary conditions (\ref{bc}) in terms of
physical quantities of the model:
in particular at the critical points (\ref{crit}) where $\phr(\xi)=S\phl(\xi)$ 
(i.e. a generalized version of the unfolded picture of the half-line), 
the boundary conditions get a very simple form
\begin{equation}
j_{\pm}(t,0,i)=\mp\sum_{j=1}^{n} \S_{ij} j_{\pm}(t,0,j)
\end{equation}
which simply fixes the splitting of the currents at the junction.
Comparing this expression with the Kirchhoff conditions (\ref{kirchV}) and 
(\ref{kirchA}), we see that at least one of two charges $Q$ and 
$\widetilde{Q}$, associated to $\rho_{+}$ and $\rho_{-}$ respectively,
is dissipated by a non trivial flow at the vertex. 
Since $\rho_{+}$ generates the electric charge for the $\psi$, 
(\ref{infinitesimal1}), we typically require the Kirchhoff's 
rule (\ref{kirchV}) to preserve 
electric charge, while $\tilde{Q}$ conservation is lost.

As for the half-line, the non trivial behavior of right-left correlators due 
to the presence of vertex, allows more non vanishing correlation functions 
with respect to the line case. 
Let us consider the two-points function for $\psi$ in the Fock representation, 
and let us focus for simplicity on the case of critical boundary 
conditions (\ref{crit}).
Imposing the Kirchhoff's rule on the charge $Q$ generated by $U(1)$,
there are four non vanishing two-points correlators:
\begin{multline}
\langle \psi_1^* (t_1,x_1,i_1) \psi_1 (t_2,x_2,i_2) \rangle  = 
\frac{z_{i_1}z_{i_2}}{2\pi} \\
\Lambda^{-[(\sigma^2+\tau^2)\delta_{i_1i_2} + 2\sigma \tau \S_{i_1i_2}]} 
[{\cal D}(vt_{12}-x_{12})]^{\sigma^2\delta_{i_1i_2}}\\ {}
[{\cal D}(vt_{12}+x_{12})]^{\tau^2\delta_{i_1i_2}} 
[{\cal D}(vt_{12}-\xt_{12}) 
{\cal D}(vt_{12}+\xt_{12})]^{\sigma \tau \S_{i_1i_2}} ,
\label{cfg1gamma} 
\end{multline}
\begin{multline}
\langle \psi_1^* (t_1,x_1,i_1) \psi_2 (t_2,x_2,i_2) \rangle  = 
\frac{z_{i_1}z_{i_2}}{2\pi} \\
\Lambda^{-[(\sigma^2+\tau^2)\S_{i_1i_2}+ 2\sigma \tau \delta_{i_1i_2}]} 
[{\cal D}(vt_{12}-\xt_{12})]^{\sigma^2 \S_{i_1i_2}}\\ {}
[{\cal D}(vt_{12}-x_{12})
{\cal D}(vt_{12}+x_{12})]^{\sigma \tau \delta_{i_1i_2}}
[{\cal D}(vt_{12}+\xt_{12})]^{\tau^2 \S_{i_1i_2}} ,  
\label{cfg2gamma} 
\end{multline}
with all normalization factors defined in appendix \ref{appA}.
All other non-vanishing correlation functions have the same form as the ones
on the half-line Eqs. (\ref{cfhl1}), (\ref{cfhl2}), and (\ref{cfhl3}) 
with only the proper wire index added.


For the charge densities one finds
\begin{multline}
\langle \rho_+ (t_1,x_1,i_1) \rho_+ (t_2,x_2,i_2) \rangle  = 
\\
\frac{-1}{(2\pi\zeta_+)^2} 
\Big\{ \left[{\cal D}^2(vt_{12}-x_{12})+{\cal D}^2(vt_{12}+x_{12})\right]
\delta_{i_1i_2} \\ +  
\left[{\cal D}^2(vt_{12}-\xt_{12})+{\cal D}^2(vt_{12}+\xt_{12})\right]
\S_{i_1i_2} \Big\}
\, , 
\label{cfgdvgamma}
\end{multline}
and for the currents
\begin{multline}
\langle j_+ (t_1,x_1,i_1) j_+ (t_2,x_2,i_2) \rangle  =\\ 
\frac{-1}{(2\pi\zeta_+)^2} 
\Big\{
\left[{\cal D}^2(vt_{12}-x_{12})+{\cal D}^2(vt_{12}+x_{12})\right]
\delta_{i_1i_2}
\\ -  
\left[{\cal D}^2(vt_{12}-\xt_{12})+{\cal D}^2(vt_{12}+\xt_{12})\right]
\S_{i_1i_2}
\Big \}\, .  
\label{cfgcvgamma}
\end{multline} 
The opposite signs in the $\delta_{i_1i_2}$ and $\S_{i_1i_2}$ contributions 
in (\ref{cfgcvgamma}) ensure the Kirchhoff's rule for $Q$.  
Analogous expressions hold for $\rho_{-}$ and $j_{-}$ up to replace in Eqs.
(\ref{cfgdvgamma}) and  (\ref{cfgcvgamma})
$(\tau+\sigma)\leftrightarrow(\tau-\sigma)$ and $\S\leftrightarrow-\S$.

If instead we impose the conservation of the charge ${\widetilde Q}$ 
we have the non-vanishing two-point correlation functions 
\begin{multline}
\langle \psi_1^* (t_1,x_1,i_1) \psi_1 (t_2,x_2,i_2) \rangle  =
\frac{z_{i_1}z_{i_2}}{2\pi}\\
\Lambda^{-[(\sigma^2+\tau^2)\delta_{i_1i_2}+2\sigma\tau \S_{i_1i_2}]} 
{}[\D(vt_{12}-x_{12})]^{\sigma^2\delta_{i_1i_2}}\\
[\D(vt_{12}+x_{12})]^{\tau^2\delta_{i_1i_2}} 
[\D(vt_{12}-\xt_{12}) \D(vt_{12}+\xt_{12})]^{\sigma \tau \S_{i_1i_2}}\, ,
\label{cfg1a}
\end{multline}
\begin{multline}
\langle \psi_1 (t_1,x_1,i_1) \psi_2 (t_2,x_2,i_2) \rangle  = 
\frac{z_{i_1}z_{i_2}}{2\pi}\\
\Lambda^{[(\sigma^2+\tau^2)\S_{i_1i_2} + 2\sigma \tau \delta_{i_1i_2}]} 
[\D(vt_{12}-x_{12})\D(vt_{12}+x_{12})]^{-\sigma \tau \delta_{i_1i_2}}\\
[\D(vt_{12}-\xt_{12})]^{-\sigma^2 \S_{i_1i_2}}
[\D(vt_{12}+\xt_{12})]^{-\tau^2 \S_{i_1i_2}}\, ,
\label{cfg2a}
\end{multline}
\begin{multline}
\langle \psi_1 (t_1,x_1,i_1) \psi_1^* (t_2,x_2,i_2) \rangle  = \\
\langle \psi_1^* (t_1,x_1,i_1) \psi_1 (t_2,x_2,i_2) \rangle \, ,\\
\shoveleft{\langle \psi_2 (t_1,x_1,i_1) \psi_1 (t_2,x_2,i_2) \rangle  =} \\
\langle \psi_2^* (t_1,x_1,i_1) \psi_1^* (t_2,x_2,i_2) \rangle =\\
\langle \psi_1^* (t_1,x_1,i_1) \psi_2^* (t_2,x_2,i_2) \rangle = \\
\langle \psi_1 (t_1,x_1,i_1) \psi_2 (t_2,x_2,i_2) \rangle\, ,
\end{multline}
and
\begin{multline}
\langle \psi_2^* (t_1,x_1,i_1) \psi_2 (t_2,x_2,i_2) \rangle  = \\
\langle \psi_2 (t_1,x_1,i_1) \psi_2^* (t_2,x_2,i_2) \rangle
= (\ref{cfg1a}) \quad {\rm with}\quad  \sigma \leftrightarrow \tau \, .
\label{cfg5a}
\end{multline}
The non conservation of the electrical charge is explicitly shown by the
presence of non-neutral correlator $\langle\psi\psi\rangle$.
The correlations for conserved density $\rho_-$ and current $j_-$ are the
same as Eqs. (\ref{cfgdvgamma}) and (\ref{cfgcvgamma}).

\section{RG flow on the junction}
\label{SecF}

\begin{figure}[t]
\includegraphics[width=\columnwidth]{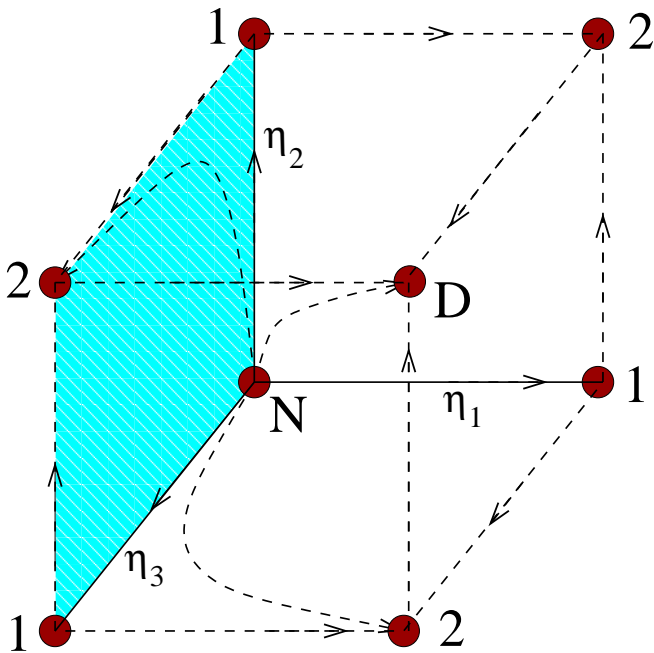} 
\caption{(Color online) RG flow diagram for a junction of three wires in the
  $\eta_{1,2,3}$ space. The fixed points are: $D$ is Dirichlet and corresponds
  to all $\eta_i=\infty$; $N$ is Neumann with $\eta_i=0$; $1$ are three fixed
  points families (depending on two parameters) with two $\eta$ vanishing and 
  one infinite; $2$ are three fixed points two-parameter families with 
  one $\eta$ zero and two infinite.
  The cyan-shaded area is the allowed region when the Kirchhoff's rule for the
  electric charge is valid.  It includes Neumann, two $p=1$ families (with
  only one parameter left free) and one $p=2$ fixed point (with no free
  parameter left).
  The arrow in the flow corresponds to the attractive case with 
  $g>1$ which gives Dirichlet as the most stable fixed point
  (without Kirchhoff) or the mixed $p=2$ (with Kirchhoff). 
  In the opposite repulsive case $g<1$, all the arrows are reversed and the most
  stable fixed point is Neumann.
}
\label{RGflow}
\end{figure}

We completely characterized the fixed-point structure for a junction with
an arbitrary number of wires $n$.
Let us recall the main features explained in the previous section and in
the appendix \ref{appcrit}. 
At the critical point, the scattering matrix can only have eigenvalues $\pm1$.
For generic $n$, the fixed points are classified in terms of the integer 
number $p$ with $0\leq p\leq n$, which is the number of eigenvalues equal 
to $-1$.
At the fixed point, the boundary couplings $\eta_i$ (with $1\leq i\leq n$) are 
zero if the corresponding eigenvalue is $+1$, infinity if the eigenvalue 
is $-1$. $p=0$ corresponds to Neumann boundary conditions on all wires, 
while $p=n$ to Dirichlet.
Other values of $p$ correspond to intermediate boundary conditions, that are
$n-p$ Neumann and $p$ Dirichlet fields in the basis $\ph_i^d$ diagonalizing 
the S-matrix. 
In Fig. \ref{RGflow} we report as a typical example the RG flow diagram
for three wires in the $\eta_i$ space. The final point of any axis is
$\eta_i=\infty$.
Let us discuss now the structure of the fixed points,
postponing the study of the stability to the following.
There are $2^3=8$ fixed points families, one Neumann, 3 points with $p=1$, 
three with $p=2$ and one Dirichlet [in the general case, there are 
$2^n$ families of which $\binom{n}{p}$ for any $p$]. 
Every critical point belongs to a continuous family with $p(n-p)$ real 
parameters that are not shown in Fig. \ref{RGflow}. 
Summarizing any critical point is identified by $p$, by the specific
eigenvalues that are $-1$ (i.e. by the axis in the figure) and by 
the $p(n-p)$ real parameters. 
The parameters specifying the fixed point in the families are the angles 
$\a_i$ reported for some examples in appendix \ref{appcrit}. 
For a given situation, the fixed point value of 
$\a_i$ is given by their initial values. This means that $\a_i$ are
marginal couplings and their values cannot be fixed only by requiring scale
invariance.

The role played by the conservation rules in this flow diagram is 
fundamental. 
To consider the most physical case, let us discuss when the electrical charge
is conserved, i.e. the Kirchhoff rule $\sum_i j_+(t,0,i)=0$ is satisfied. 
The first effect is to fix to zero one (arbitrary) $\eta_i$, constraining 
the system on the shadow area in Fig. \ref{RGflow} so that Dirichlet boundary
conditions are ruled out for the problem. 
Also the number of real parameters characterizing the $p$-fixed points is
largely reduced. For three wires, the point with $p=1$ becomes a one-parameter
family, while the point with $p=2$ becomes an isolated fixed point. 
Details for the general case are in the appendix. 
Needless to say that imposing the conservation of the
$\widetilde{U}(1)$ charge, results in fixing one of the $\eta_i$ to 
$\infty$ and similarly reduced the number of real parameters available for
each fixed point.

We briefly discuss our terminology for the fixed points, in order to make 
the comparison with other papers as simple as possible.
The fixed points with $p=2$ (``2'' in Fig. \ref{RGflow}) is the mixed fixed 
point found by Nayak et. al \cite{nfll-99} and called $D$ (or $D_P$) in 
Ref. \onlinecite{coa-03} because of the $n-1$ Dirichlet boundary conditions
(there $n=3$)
on the neutral modes (but this point is obviously different from our $D$).
The family with $p=1$ in Fig. \ref{RGflow} depends on a continuous real 
parameter $\a$, as shown in Eq. (\ref{famp1}), and it has been first found in 
Ref. \onlinecite{bms-07}. 
Note that it is not symmetrical under wire permutations.
There are three special values of $\a$: for $\a=-1,0,\infty$ the 
$S$-matrix breaks  into a $1\times 1$ and a $2\times2$ blocks. 
The $1\times 1$ block is a wire decoupled from the other two that form a 
purely transmitting $n=2$ junction (the same can be verified for higher $n$, 
changing the $\alpha$'s we can decouple any wire).
For these special values of $\a$, the fixed points were also found by Chamon 
et al. \cite{coa-03} that called them asymmetrical $D_A$.
Other values of $\a$ interpolate continuously between these three.
Finally it is worth commenting that the Dirichlet fixed point (D in 
Fig. \ref{RGflow}) physically corresponds to $n$ wires with an end inserted 
into a large superconductor. 
In fact, the S-matrix $S=-\II$ gives conductance $\G=2\II$ corresponding to 
Andreev reflection in all wires (i.e. sending a particle one gets an hole out).
This is a different problem from a junction of wires (even superconducting), 
because the large superconductor breaks the $U(1)$ charge 
conservation \cite{coa-03,dr-08}.

Now we know the fixed-point structure, but what is the relative stability? 
Which fixed point describe the universal low energy behavior?
There are several equivalent ways to tackle this question. 
The more natural one, as done elsewhere \cite{nfll-99,coa-03,hc-08}, relies on
calculating the scaling dimension of the perturbing operator at a
given fixed point. 
Since our problem can be thought as $n$ independent half-lines with $n-p$ 
Neumann boundary conditions and $p$ Dirichlet ones, the problem is just
equivalent to understand the stability of Neumann or Dirichlet against a Robin
term as in the Hamiltonian (\ref{Hambou}). 
This is a standard problem.
In the bosonic theory, the flow can be followed exactly from Eq. (\ref{Sflow})
of the off-critical $S$-matrix.
The Neumann fixed point is always unstable, 
while Dirichlet is stable (or mixed if Kirchhoff is imposed on the electrical 
charge).
However, as well known, considering the fermionic theory changes this scenario 
because of the Klein factors. 
In boundary conformal field theory, the stability conditions are just read
from the boundary dimensions $d_b$ appearing in the
two-points correlation functions reported above.
At the Neumann BC we have that the dimension is
$\sigma\tau=(\zeta_+^2-\zeta_-^2)/4$ that is greater than zero for $g_+>g_-$,
i.e. for repulsive anyonic interaction, giving a stable Neumann.
Oppositely at the Dirichlet BC the boundary dimension is 
$-\sigma\tau$ that it is stable in the complementary attractive case.
Since there are no other fixed points in the RG diagrams, this analysis fixes
all the RG flow.
Note that for free anyons (and in particular fermions) $\eta$ is marginal in
this approach. In any given anyonic/fermionic model the actual stable fixed 
point will be determined by the higher order terms in $\eta$ neglected in our
approach.

These results can be confirmed on the basis of the following argument based 
on the so called $g$-theorem \cite{al-91,fk-04}. 
For a one-dimensional critical system with a boundary, 
it is known that the boundary contribution to the entropy $\ln g$ ($g$ is the 
so called ``universal non integer ground state degeneracy''\cite{al-91}) 
decreases along the renormalization group flow.
We can easily calculate the value of the effective-potential 
$V_{\rm eff}= g_+\rho_++g_-\rho_-$ for the off critical model for any $\eta$.
Subtracting the divergent contribution of the bulk to make this expectation
value finite, we get on each wire
\begin{equation}
  \varepsilon(x,i)=\langle V_{\rm eff}(t,x,i)\rangle=\Omega
\int_{-\infty}^{+\infty}\frac{dk}{2\pi}|k|e^{2ikx} \S_{ii}(k)
\label{energy1}
\end{equation}
where 
\begin{equation}
\Omega=\left(\frac{g_-\zeta_{-}^{2}-g_+\zeta_{+}^{2}}{2\pi\kappa^2}\right)\,,
\end{equation}
fully encodes the bulk interactions effect. 
In particular, when $g_{+}=g_{-}$ it vanishes and changes sign, giving the 
correct stability scenario.

In fact, we can rewrite (\ref{energy1}) in terms of the potential
$\varepsilon_{\eta_{j}}(x)$ for disjointed half-line with the boundary 
condition (\ref{bchlg})
\begin{equation}
\varepsilon(x,i)=\sum_{j}^{n}|\U_{ji}|^2 \varepsilon_{\eta_{j}}(x)\,,
\end{equation}
with
\begin{equation}
\varepsilon_{\eta}(x)=-\frac{\Omega}{4x^2}
\left[1-4(x\eta)-8(x\eta)^2 e^{2x\eta}\rm{Ei}(-2x\eta)\right]\,.
\label{energy2}
\end{equation}
The function 
\begin{equation}
s(x)=-4x^2\sum_{i=1}^{n}  \varepsilon(x,i)=
-4x^2\sum_{j=1}^{n}\varepsilon_{\eta_{j}}(x)\,,
\end{equation}
collects the contribution of all the wires. It is a monotonous function
with fixed points at $\eta=0,\infty$ in agreement with the $g$ theorem. 
The stability of the fixed points and the direction of the flow are 
just given by the sign of $\Omega$ and agrees with the previous analysis.

\section{Conclusions}
\label{SecC}

In this paper we presented a systematic study of the critical properties of
$n$ anyonic Luttinger wires jointed in a single vertex. 
Imposing the boundary conditions (\ref{bc}) at the junction directly on 
bosonized fields allowed us to describe completely the RG flow diagram for 
any $n$. As a typical example the RG flow for $n=3$ is depicted in Fig. 
\ref{RGflow} where the main features of the various fixed points are 
discussed in the text. 

At this point it is worth comparing our findings with the literature.
For two wires, our results are a simple anyonic generalization of  
the well-known ones by Kane and Fisher \cite{kf-92} for fermions that are
reproduced for $\k=1$.
For $n=3$, as we said in the introduction the literature is enormous.
The boundary conditions we used are equivalent to those of the 
``auxiliary model'' of Nayak et al.\cite{nfll-99} for $g\neq1$
[in fact, expanding the exponential defining the auxiliary model \cite{nfll-99}
and keeping only up to the quadratic terms, neglecting irrelevant higher
orders, we arrive to the Hamiltonian (\ref{Hambou}) where the symmetry of the
boundary terms is just the Kirchhoff's rule]. 
We predict two possible stable fixed points: Neumann and mixed.
Neumann is well known, it has zero conductance and in this setting 
it is stable for all repulsive interactions, i.e. $g<1$. 
The mixed fixed point has been found for the first time by 
Nayak et al. \cite{nfll-99} and it is specific of the junctions. 
It has enhanced conductance $G/G_{\rm line}=4/3$ and we found it is stable 
for all attractive interactions $g>1$ as in Ref. \onlinecite{nfll-99}. 
Everything agrees with the auxiliary model, but not with the ``standard
model'' defined by the boundary condition (\ref{blagrangian}), 
that is known to be different\cite{nfll-99}.
In fact in the standard model, the Neumann fixed point is stable only for
$g<1/3$ while the mixed one only for $g>9$. In the other regimes with $1/3<g<9$,
new fixed points appear that {\it cannot} be present in our 
approach \cite{nfll-99,coa-03}. 
Our setting however presents a great advantage: it is simpler for generic 
$n$ and more efficient in describing the off-critical properties of the system.
In fact we provide for the first time the critical behavior for all $n$.
We found for $g<1$ a Neumann stable fixed point (with zero conductance) 
and for $g>1$ a mixed fixed point with conductance $G/G_{\rm line}= 2(n-1)/n$. 
We also find other fixed points (described in the appendix \ref{appcrit})
that however have at least one direction of instability in the $\eta_i$ space
and so they are multicritical points, in the sense that some other 
constraints must be imposed to reach them.
Clearly we expect that the standard model for $n\geq3$ will have some fixed
points not found here, as for the case $n=3$. 
A part from the per se interest of the model, the fixed points we found are
relevant for the standard model as well. 
In fact, it is easy to generalize to any $n$ the strong and weak boundary 
coupling (i.e. our $\eta$) calculations of 
Refs. \onlinecite{nfll-99,coa-03} to show that for small enough $g$ the
relevant fixed point is Neumann and for large enough is the mixed one.
However, which fixed point governs the dynamics when none of these two is 
stable is not accessible to our approach. 
For $n=4$, two fixed points derived in the Appendix \ref{appcrit}
have been recently
found to describe the scattering matrix for a proposed experiment to 
detect the helical nature of the edge states in quantum Hall 
systems \cite{hkc-08}.

We mention that we also characterized the junction in the absence of the
Kirchhoff's rule for the electric charge. It is of particular relevance
considering the case when relaxing the conservation of the electrical charge 
and imposing the conservation of the dual one ${\widetilde U}(1)$.
In this case the more stable fixed point is always Dirichlet with uncommon
points like the mixed one representing multicritical points.

There are two generalizations of the model considered here that should be 
easily accessible to a similar analysis. 
First of all one can consider fermions with spin (and even multispecies
anyons) as done elsewhere with fermionic boundary conditions \cite{hc-08}.
In this way one can understand which fixed points are present also with bosonic
boundary conditions. 
The other generalization is relaxing the symmetry for time reversal to allow a
non vanishing flux at the junction \cite{coa-03}.

We close this paper on a more speculative level. 
In recent times there has been an increasing interest in quantifying the
entanglement in extended quantum systems (see e.g. \onlinecite{rev} as 
reviews). 
Among the various measures, the so-called entanglement entropy
has by far been the most studied. 
By partitioning an extended quantum system into two blocks,
the entanglement entropy is defined as the von Neumann
entropy of the reduced density matrix $\rho_A$ of one of the two blocks. 
This procedure requires an arbitrary division of the system in two parts. 
In the junction problem studied here the system is automatically divided in
parts and it would be very interesting to understand the amount of entanglement
between the various wires. 
The analysis of some models on the line with one defect \cite{pes} 
(i.e. $n=2$ in the language of this paper) 
showed that the entanglement entropy is not always only
dependent on the central charge of the bulk theory (as maybe naively expected).
The natural question is whether the conformal field theory formalism that has
been successfully applied to the bulk and boundary case \cite{ee} can be
generalized to the junction.
Furthermore, if we would be able to solve the non-equilibrium problem with
changing the boundary condition (e.g. suddenly adding or removing the 
junctions, as done for $n=2$ in Ref. \onlinecite{cc-07}), one can think of 
using the junction as an {\it entanglement meter} following the recent 
proposal based on quantum noise measurement \cite{kl-08}.

\acknowledgments
We thank Claudio Chamon for fruitful comments on a first version of the 
manuscript and for useful discussions. 
PC benefited of a travel grant from ESF (INSTANS activity).

\appendix

\section{Bosonization and quantization of the TL model}
\label{appA}

\subsection{The line}

The basic tool for quantizing the system, described by the Eqs. (\ref{eqm1}), 
is the algebra ${\cal A}$ generated by the bosonic annihilation $a(k)$ and 
creation $a^*(k)$ operators satisfying 
\begin{eqnarray} 
&&\label{ccr} [a(k)\, ,\, a(p)] = [a^* (k)\, ,\, a^* (p)] = 0 \,, \\
&&[a(k)\, ,\, a^* (p)] = 4\pi |k^{-1}|_\Lambda   \delta (k-p) \, ,
\end{eqnarray} 
where the normalization can be fixed such that
\begin{equation} 
|k^{-1}|_{\Lambda } = \frac{\rd}{\rd k}
\left[ \theta(k) \ln\frac{k\, \e^{\gamma_E}}{\Lambda }\right ] \, .
\label{distr2}
\end{equation} 
The derivative here is understood in the sense of distributions, 
$\gamma_E$ is Euler's constant and $\Lambda>0$ is a free parameter with 
dimension of mass having a well-known infrared origin. 
It is useful to introduce
\begin{multline} 
u(\Lambda \xi )\equiv \int_{0}^{\infty} \frac{\rd k}{\pi }|k^{-1}|_{\Lambda }  \e^{-\ri k\xi} = 
-\frac{1}{\pi} \ln (\Lambda |\xi|) -\frac{\ri}{2}\varepsilon (\xi) \\= 
-\frac{1}{\pi} \ln (\ri \Lambda \xi + \epsilon )\, , \qquad \epsilon > 0\, . 
\label{log}
\end{multline} 
The left and right chiral fields are defined by 
\begin{eqnarray}
 \phr (\xi) &=&
{} \int_0^\infty \mis \left [a^* (k) \e^{\ri k \xi} +
a(k)\e^{-\ri k \xi } \right ] \, , \\
\phl (\xi) &=&
{} \int_0^{\infty }\mis \left [a^\ast (-k) \e^{\ri k \xi} +
a(-k) \e^{-\ri k \xi}\right ]  , 
\end{eqnarray}
and obey the commutation relations
\begin{equation}
[\phr (\xi _1 )\, ,\, \phr (\xi _2 )] =
[\phl (\xi _1 )\, ,\, \phl (\xi _2 )] =
-\ri \varepsilon (\xi _{12}) \, , 
\label{comm1}
\end{equation}
\begin{equation}
[\phr (\xi _1 )\, ,\, \phl (\xi _2 )] = 0 \, , 
\label{comm2}
\end{equation}
and have the correlations
\begin{equation}
\langle\ph_R (\xi_1) \ph_R(\xi_2)\rangle=
\langle\ph_L (\xi_1) \ph_L(\xi_2)\rangle=
u(\Lambda \xi_{12})\,,
\label{LLRR}
\end{equation}
with $\xi_{12}=\xi_1-\xi_2$ and obviously
$\langle\ph_R (\xi_1) \ph_L(\xi_2)\rangle=0$.

Defining the chiral charges by 
\begin{equation}
Q_{Z}= \frac{1}{4} \int_{-\infty}^\infty \rd \xi \, (\der \phz) (\xi) \, , \qquad Z=R,L\, , 
\label{chq}
\end{equation}
one gets 
\begin{eqnarray}
\,[Q_R, \phr(\xi)] &=& [Q_L, \phl(\xi)]= -{\ri}/{2} \,, \nonumber\\
\,[Q_R, \phl(\xi)] &=& [Q_L, \phr(\xi)]= [Q_R, Q_L]=0.
\label{chqcomm}
\end{eqnarray} 
It is worth mentioning that all previous the commutation relations 
are invariant under the {\it duality} transformation 
\begin{equation} 
\phr (\xi) \mapsto \phr (\xi)\, , \qquad \phl (\xi) \mapsto -\phl (\xi)\, ,
\label{dual1}
\end{equation} 
which define the T-duality in string theory. 

At this point we are ready to introduce a family of vertex operators 
parametrized by two real variables $\sigma$ and $\tau$ defined by 
\begin{equation}  
A(t,x) =  z \e^{\ri \sqrt {\pi}(\tau Q_{{}_R} -\sigma Q_{{}_L})} 
:\e^{\ri \sqrt {\pi} \left [\sigma \phr (vt-x) + \tau\phl (vt+x)\right ]}:,
\label{vertex1}
\end{equation}
with 
\begin{equation}
z=(2\pi)^{-1/2}\Lambda^{(\sigma^2 + \tau^2)/2}\, ,
\label{zl}
\end{equation} 
where $: \cdots :$ denotes the normal product in ${\cal A}$ and $v$ is some 
velocity to be determined by consistency. 
From Eqs. (\ref{psi1}) and (\ref{psi2}) the fields $\psi_1$ and $\psi_2$ are 
vertex operators with interchanged $\sigma$ and $\tau$, with a normalization 
constant given by Eq. (\ref{zl}). The factor 
$\e^{\ri \sqrt {\pi}(\tau Q_{{}_R} -\sigma Q_{{}_L})}$ is included in the
definition (\ref{zl}) to ensure canonical anionic commutation relation
between $\psi_{1,2}$ without introducing Klein factors that will be important
only for the fields on different wires.

The following identity is useful in determining the exchange properties of 
the vertex operators and so all correlation functions 
\begin{multline}
A^*(t,x_1) A(t,x_2) =  
|x_{12}|^{-(\sigma^2 + \tau^2)} 
\e^{-\ri \frac{\pi}{2}(\tau^2-\sigma^2) \varepsilon(x_{12})}\\ 
:\e^{\ri \sqrt {\pi} \left [\sigma \phr (vt-x_2) - \sigma \phr (vt-x_1) + \tau \phl (vt+x_2)-\tau \phl (vt+x_1)\right ]}:,\, 
\label{identity1}
\end{multline} 
where $x_{12}\equiv x_1-x_2$. 

The normalization of the charge densities $\rho_\pm$
is fixed by requiring that they generate the 
transformations (\ref{A}) and  (\ref{V}) in infinitesimal form, namely
\begin{align}
 [\rho_+(t,x_1),\psi_\a (t,x_2)]&=&-
\delta(x_{12})\psi_\a(t,x_2)\,, 
\label{infinitesimal1} \\
{}[\rho_-(t,x_1),\psi_\a (t,x_2)] &=&- 
(-1)^{\a} \delta(x_{12}) \psi_\a(t,x_2)\,.
\label{infinitesimal2}
\end{align}

\subsection{The half-line}

In the main text, we stressed that on the half line right and left modes couple
and have non trivial commutation relations given by Eq. (\ref{RLhl}).
This gives rise to few changes to the relations valid on the full line.
The vertex operator is always defined by Eq. (\ref{vertex1}), but the
normalization constant is affected by the boundary \cite{lm-98}:
\begin{equation}
z=
\left\{
\begin{array}{ll}
(2\pi)^{-1/2}\Lambda^{(\sigma+\tau)^2/2}\,,\; & \eta =0\,;\\
(2\pi)^{-1/2}\Lambda^{(\sigma-\tau)^2/2}\,,\; & 0<\eta \leq\infty\,.
\end{array}
\right.
\label{z2}
\end{equation}

The right-left coupling also affects the correlation functions of the field
$\ph$. In fact, while the right-right and left-left correlators are still 
given by Eq. (\ref{LLRR}), the mixed ones are
\begin{equation}
\langle\phr(\xi_{1})\phl(\xi_{2})\rangle=
\left\{
\begin{array}{ll}
u(\Lambda\xi_{12})& \eta=0,\\
-u(\Lambda\xi_{12})& \eta=\infty,\\
-u(\Lambda\xi_{12})-v_{-}(\Lambda\xi_{12})&  0<\eta<\infty,
\end{array}
\right.
\label{crl}
\end{equation}
\begin{equation}
\langle\phl(\xi_{1})\phr(\xi_{2})\rangle=
\left\{
\begin{array}{ll}
u(\Lambda\xi_{12})& \eta=0,\\
-u(\Lambda\xi_{12})& \eta=\infty,\\
-u(\Lambda\xi_{12})-v_{+}(\Lambda\xi_{12})&  0<\eta<\infty,
\end{array}
\right.
\label{clr}
\end{equation}
where the ``boundary propagator'' is 
\begin{equation}
v_{\pm}(\xi)=\frac{2}{\pi}e^{-\xi}{\rm Ei}(\xi\pm\ri\epsilon),
\end{equation}
and ${\rm Ei}(x)=\int_x^\infty dz e^{-z}/z$ is the exponential integral 
function, that at small $x$ has the right logarithm expansion.
Note that in the above formulas for mixed correlators $\xi_1=vt_1-x_1$
and $\xi_2=vt_2+x_2$ or viceversa, thus $\xi_{12}=v t_{12}\mp \xt_{12}$, 
with the sign depending on the correlator if it is right-left or left-right
respectively.

\subsection{The junction}

For the theory on the star graph, all the relevant commutation relations and
correlators of the fields follow from those on the half line after performing
the linear transformation $\U$ in Eq. (\ref{d1}). In fact all the fields
$\ph^d$ are just delocalized fields satisfying the proper
boundary conditions reported above with different $\eta_i$ for each mode.
Thus, comparing with the half line equations (\ref{RLhl}), it is 
straightforward to derive the commutation relations for the right and 
left movers on the wires
\begin{multline}
[\ph_{i_{1},R}(\xi_{1}),\ph_{i_{i2},R}(\xi_{2})]= 
[\ph_{i_{1},L}(\xi_{1}),\ph_{i_{2},L}(\xi_{2})]\\
=-\ri\epsilon(\xi_{12})\delta_{i_{1}i_{2}}\,,
\label{RRLLgamma}
\end{multline}
\begin{equation}
[\ph_{i_{1},R}(\xi_{1}),\ph_{i_{2},L}(\xi_{2})]=
\U^{-1}_{i_{1}k_{1}}\U_{l_{2}i_{2}}
[\ph^{d}_{k_{1},R}(\xi_{1}),\ph^{d}_{l_{2},L}(\xi_{2})]\,,
\label{RLgamma}
\end{equation}
where $\ph_{R,L}^{d}(\xi)=\U\ph_{R,L}(\xi)$ and
\begin{multline}
[\ph^d_{i_{1},R}(\xi_{1}),\ph^d_{i_{2},L}(\xi_{2})]=\\
 \left\{
 \begin{array}{ll}
 -\ri\epsilon(\xi_{12})\delta_{i_{1}i_{2}}\,,& \eta_{i_{1}}=0\,;\\
 \ri\epsilon(\xi_{12})\delta_{i_{1}i_{2}}\,, & \eta_{i_1}=\infty\,;\\
\left[\ri\epsilon(\xi_{12})-4\ri\theta(\xi_{12})e^{-\eta_{i_{1}}\xi_{12}}\right] \delta_{i_{1}i_{2}}\,, &
 0<\eta_{i_{1}}<\infty\,.
 \end{array}
 \right.
\end{multline}
The mixed commutator (\ref{RLgamma}) simplifies greatly for critical boundary 
conditions 
\begin{equation}
[\ph_{i_{1},R}(\xi_{1}),\ph_{i_{2},L}(\xi_{2},i_{2})]= 
-\ri\epsilon(\xi_{12})\S_{i_{1}i_{2}}\,.
\end{equation}
Note that at spacelike distances where $vt_{12}-\xt_{12}<0$, 
the commutators (\ref{RRLLgamma}) and (\ref{RLgamma}) behave
as if the scattering matrix were replaced by the critical one obtained  
in the infrared limit $\Lambda\rightarrow\infty$ 
or equivalently $k\rightarrow0$ 
\begin{equation}
[\ph_{i_{1},R}(\xi_{1}),\ph_{i_{2},L}(\xi_{2})]
_{|_{(v^2t_{12}^{2}-x_{12}^2<0)}}=-\ri \S_{i_{1}i_{2}}(0)\,.
\label{RLspace}
\end{equation}
This simply means that $\ph_{R,L}$ has the same properties of locality 
than its infrared limit. 

The last complication on the star graph arises in the definition of the anyonic
fields $\psi_{1,2}$. To have the correct commutation relation they must be 
defined according to
\begin{multline}  
\psi_1(t,x,i) = z_i\, \eta_i \, 
\e^{\ri \sqrt{\pi}\left (\tau Q_{i,R} -\sigma Q_{i,L}\right )} \\ \times
:\e^{\ri \sqrt{\pi}\left [\sigma \ph_{i,R}(vt-x) + 
\tau \ph_{i,L}(vt+x)\right]}: \, , \\
\shoveleft{\psi_2(t,x,i) = z_i\, \eta_i \, 
\e^{\ri \sqrt{\pi}\left (\sigma Q_{i,R} -\tau Q_{i,L}\right )} }\\ \times
:\e^{\ri \sqrt{\pi}\left [\tau \ph_{i,R}(vt-x) + 
\sigma \ph_{i,L}(vt+x)\right]}: \, , 
\end{multline} 
where $z_i$ are fixed to
\begin{equation}
z_{i}=(2\pi)^{-1/2}\Lambda^{[(\sigma^2+\tau^2)+2\sigma\tau \S_{ii}(0)]/2}\,,
\end{equation}
and $\eta_i$ are the anyonic Klein factors needed to ensure the correct
commutation of anyon fields on different edges
\begin{equation}
\psi(t,x_i,i) \psi(t,x_j,j)=\e^{-i\pi \k\eps_{ij}}\psi(t,x_j,j)\psi(t,x_i,i)\,,
\end{equation}
where $\eps_{ij}=\eps(i-j)$. 
It is straightforward to build them for example in terms of
the auxiliary Majorana algebra $[c_i,c_j]=\ri \k\eps_{ij}$ and $c_i^*=c_i$
resulting in $\eta_i=:\e^{\pi\ri c_i}:$.
These factors are of fundamental importance when considering as junction
condition Eq. (\ref{blagrangian}), because it is written in terms of anyonic 
degrees of freedom. 
Oppositely, because the junction condition we use is written in terms
of currents that only get (re)normalized by the statistics, they are
inessential. For this reason we do not discuss them further,   
remanding the interested readers to the complete treatment presented 
in Ref. \onlinecite{gdms-02} and in the appendix E of \onlinecite{coa-03}.

\section{Critical points} 
\label{appcrit}

By scale invariance any critical point is associated with a $k$-independent 
$S$-matrix satisfying unitarity (\ref{unit1}), Hermitian 
analyticity (\ref{Ha}) and time-reversal invariance (\ref{tr2}), i.e. 
\begin{equation}
\S^*=\S^{-1}\, ,\qquad \S^*=\S\, , \qquad \S^t=\S\,. 
\label{sc3}
\end{equation} 
The classification of these $S$-matrices is now a simple 
matter. Indeed, one can easily deduce from (\ref{sc3}) that the eigenvalues of 
$\S$ are $\pm 1$. Let us denote by $p$ the number of eigenvalues $-1$. 
The values $p=0$ and $p=n$ correspond to the familiar 
Neumann ($\S_N=\II$) and Dirichlet ($\S_D=-\II$) boundary 
conditions respectively. A richer structure appears for $0<p<n$. In that case  
the $S$-matrices satisfying (\ref{sc3}) depend on $p(n-p)\geq 1$ parameters, 
giving raise to whole families of critical points \cite{bm-06,bms-07}. 
Let us describe them explicitly for $n=2,3,4$.

The only possibility for $n=2$ is $p=1$, leading to 
the one-parameter family \cite{Bachas:2001vj,ms-05} 
\begin{equation}
\S = \frac{1}{1+\a^2}
\left(\begin{array}{cc}
\a^2-1&-2\a\\ 
-2\a&1-\a^2
\end{array}\right)\, , 
\label{scnk}
\end{equation} 
with $\a$ a real number. 
For $\a = -1$ one has full transmission and no reflection, which 
corresponds to the theory on the whole line. This is an example of exceptional 
boundary conditions already mentioned \cite{foot}. 
It is only the only $S$ matrix in the family satisfying Kirchhoff's rule for
the electric charge. 
Oppositely, $\a=1$ is the only matrix satisfying Kirchhoff's rule for 
the ${\widetilde U}(1)$ charge, as predicted by duality.

In the case $n=3$ one has two possibilities: $p=2$ and $p=1$. 
In both cases one has a family with two real parameters $\a_{1,2}$:
\begin{multline} 
\S_2(\a_1,\a_2) =\frac{1}{1+\a_1^2 +\a_2^2} \times \\
\left(\begin{array}{ccc}
\a_1^2-\a_2^2 -1&2\a_1 \a_2 &2\a_1 \\ 
2\a_1\a_2&-\a_1^2 +\a_2^2 -1&2\a_2\\
2\a_1&2\a_2&1-\a_1^2-\a_2^2  
\end{array}\right),
\label{famp2}
\end{multline} 
and 
\begin{equation} 
\S_1(\a_1,\a_2) = - \S_2(\a_1,\a_2) \, .   
\label{famp1}
\end{equation} 
For generic values of the parameters these $S$-matrices violate both 
$U(1)$ and ${\widetilde U}(1)$. Preserving $U(1)$, one must impose 
(\ref{kirchV}) on (\ref{famp2}). This implies 
$\a_1=\a_2=1$, leading to the isolated critical point
\begin{equation} 
\S_2 =\frac{1}{3} 
\left(\begin{array}{ccc}
-1&2&2\\ 
2&-1&2\\
2&2&-1  
\end{array}\right) \, ,  
\label{sc2}
\end{equation}
which is invariant under edge permutations.  
From (\ref{famp1}) one obtains instead $\a_2 = -(1+\a_1)$. 
Therefore, setting $\a \equiv \a_2$, one has in this case 
the one-parameter family of critical points  
\begin{equation} 
\S_1 =\frac{1}{1+\a +\a^2} 
\left(\begin{array}{ccc}
-\a&\a(\a +1)&1+\a\\ 
\a(\a +1)&\a +1&-\a\\
\a +1&-\a&\a(\a +1)  
\end{array}\right)\, , 
\label{sc4}
\end{equation} 
which is {\it not} invariant under edge permutations for generic $\a$. 
Summarizing, the critical points which respect $U(1)$ are $\S_0=\II_3$, 
(\ref{sc2}), and (\ref{sc4}). 
The matrix (\ref{sc2}) has been discovered by means of RG techniques by 
Nayak et al. \cite{nfll-99}. 
The family (\ref{sc4}) appeared for the first time in \onlinecite{bms-07}.  

If one wants to preserve ${\widetilde U}(1)$, one must require 
(\ref{kirchA}). One is left 
therefore with $\S_3=-\II_3$,    
\begin{equation} 
\S_2 =-\frac{1}{1+\a +\a^2} 
\left(\begin{array}{ccc}
-\a&\a(\a+1)&1+\a\\ 
\a(\a +1)&\a +1&-\a\\
\a +1&-\a& \a(\a +1)  
\end{array}
\right),
\label{sc4a}
\end{equation} 
and 
\begin{equation} 
\S_1 =-\frac{1}{3} 
\left(\begin{array}{ccc}
-1&2&2\\ 
2&-1&2\\
2&2&-1  
\end{array}\right) \, ,  
\label{sc2a}
\end{equation} 
as predicted by duality. 

For $n=4$ the general matrices satisfying all the constraints (\ref{sc3}) 
are too large to be reported here. 
Thus we only give the critical points for $n=4$ 
satisfying the Kirchhoff's rule Eq. (\ref{kirchV}) for the electrical current
[the analogous ones with the Kirchhoff's rule Eq. (\ref{kirchV}) are just $-\S$
because of duality]. 
Besides $\S_0=\II_4$ corresponding to $p=0$, one has: 
 
(i) for $p=1$ the $\S$-matrix depends on two real parameters $\a_{1,2}$ 
and results to be
\begin{eqnarray} 
\S_{11}&=& \frac{1}{\Delta_1}(\a_1 +\a_1 ^2+\a_2 +\a_1  \a_2 +\a_2 ^2 )\, ,\nonumber \\
\S_{22}&=& \frac{1}{\Delta_1}(1+\a_1 +\a_1 ^2+\a_2 +\a_1  \a_2 )\, ,\nonumber \\
\S_{33}&=& \frac{1}{\Delta_1}(1+\a_1 +\a_2 +\a_1  \a_2 +\a_2^2 )\, ,\nonumber \\
\S_{44}&=& -\frac{1}{\Delta_1}(\a_1+\a_2+\a_1 \a_2)\, ,\nonumber
\end{eqnarray}
\begin{eqnarray}
\S_{12}&=&-\frac{1}{\Delta_1}\a_2\, ,\quad \quad \; \; 
\S_{13}=-\frac{1}{\Delta_1}\a_1\, ,\nonumber\\
\S_{14}&=&\frac{1}{\Delta_1}(1+\a_1+\a_2)\, ,
\nonumber \\
\S_{23}&=&-\frac{1}{\Delta_1}\a_1 \a_2\, ,\quad \; \, 
\S_{24}=\frac{1}{\Delta_1}\a_2(1+\a_1+\a_2)\, ,\nonumber\\
\S_{34}&=&\frac{1}{\Delta_1}\a_1(1+\a_1+\a_2)\, , 
\nonumber 
\end{eqnarray}
\medskip 
with $\Delta_1=1+\a_1+\a_1 ^2+\a_2 +\a_1 \a_2 +\a_2 ^2$. The remaining entries 
are recovered by symmetry. Note that this matrix is not invariant under edge
permutations.

(ii) for $p=2$ the $S$-matrix still depends on two real parameters: 
\begin{eqnarray} 
\S_{11} &=& \frac{1}{\Delta_2}[3 \a_1 ^2+2 \a_1  (1-\a_2 )-(1+\a_2 )^2 ]\, ,\nonumber \\
\S_{22} &=& \frac{1}{\Delta_2}[-1-\a_1 ^2+2 \a_2 +3 \a_2 ^2-2 \a_1 (1+\a_2 )]\, ,\nonumber \\
\S_{33} &=& \frac{1}{\Delta_2}[3-\a_1^2+2\a_2-\a_2^2+2\a_1(1+\a_2)]\, ,\nonumber \\
\S_{44} &=& -\frac{1}{\Delta_2}[\a_1 ^2+2 \a_1  (1-\a_2 )+(1+\a_2 )^2]\, ,\nonumber \\
\nonumber
\end{eqnarray} 
\begin{eqnarray}
\S_{12}&=&\frac{2}{\Delta_2}(1+\a_1 +\a_2 +2 \a_1 \a_2 )\,,
\nonumber\\
\S_{13}&=&\frac{2}{\Delta_2}[\a_2 (1+\a_2 )-\a_1  (2+\a_2 )]\,,
\nonumber \\
\S_{14}&=&\frac{2}{\Delta_2}(1+\a_1 -\a_1  \a_2 +\a_2 ^2)\,,
\nonumber\\
\S_{23}&=&\frac{2}{\Delta_2}(\a_1+\a_1^2-2\a_2-\a_1\a_2)\,,
\nonumber \\
\S_{24}&=&\frac{2}{\Delta_2}(1+\a_1 ^2+\a_2 -\a_1  \a_2 )\,,
\nonumber\\
\S_{34}&=&\frac{2}{\Delta_2}(\a_1 +\a_1 ^2+\a_2 +\a_2 ^2)\, , 
\nonumber 
\end{eqnarray}
where 
$\Delta_2={3+3\a_1 ^2+2\a_1  (1-\a_2)+2 \a_2 +3 \a_2 ^2}$. 
Also this matrix is not invariant under edge permutations.

(iii) for $p=3$ the we have only an isolated $S$-matrix: 
\begin{equation} 
\S =\frac{1}{4} 
\left(\begin{array}{cccc}
-2&2&2&2\\ 
2&-2&2&2\\
2&2&-2&2\\
2&2&2&-2
\end{array}\right) \, , 
\end{equation}
which is invariant under edge permutation.
This is the analogous for four wires of the Nayak et al. result
\cite{nfll-99}. 

Recently, the $p=1$ matrix with $\a_1=1$ and $\a_2=-1$ and the $p=3$ matrix 
has been found to describe the scattering matrix for a proposed experiment to 
detect the helical nature of the edge states in quantum Hall systems.

We conclude this appendix with the matrix with $p=n-1$ for general $n$
satisfying the electric Kirchhoff rule:
\begin{equation} 
\S =\frac{1}{n}
\left(\begin{array}{ccccc}
(2-n)&2&2& \cdots &2\\ 
2&(2-n)&2& \cdots &2\\
\vdots&\vdots&\vdots&\cdots&\vdots\\ 
2&2&2&\cdots&(2-n)   
\end{array}\right)\,.
\end{equation}
since it is the most stable in the
RG phase diagram as shown in the text.
This is the only matrix which is invariant under wire permutations
(i.e. that has all diagonal elements equal and non-diagonal as well),
satisfying the Kirchhoff's rule and with all non-vanishing entries.


\begin{thebibliography}{99}
\bibitem{foot}
Excluding some exceptional boundary conditions in graphs with even number 
$n=2m$ of edges for which the system behaves as a bunch of $m$ independent 
lines.
\end{thebibliography}
\end{document}